# Unusual Resistance Hysteresis in *n*-Layer Graphene Field Effect Transistors Fabricated on Ferroelectric Pb(Zr$_{0.2}$Ti$_{0.8}$)O$_3$

X. Hong,[1] J. Hoffman,[2] A. Posadas,[2] K. Zou,[1] C. H. Ahn,[2] J. Zhu[1]

[1] Department of Physics, The Pennsylvania State University, University Park, PA 16802

[2] Department of Applied Physics, Yale University, New Haven, CT 06520

**Abstract:**

We have fabricated *n*-layer graphene field effect transistors on epitaxial ferroelectric Pb(Zr$_{0.2}$Ti$_{0.8}$)O$_3$ (PZT) thin films. At low gate voltages, PZT behaves as a high-$\kappa$ dielectric with $\kappa$ up to 100. An unusual resistance hysteresis occurs in gate sweeps at high voltages, with its direction opposite to that expected from the polarization switching of PZT. The relaxation of the metastable state is thermally activated, with an activation barrier of 50-110 meV and a time constant of 6 hours at 300 K. We attribute its origin to the slow dissociation/recombination dynamics of water molecules adsorbed at the graphene-PZT interface. This robust hysteresis can potentially be used to construct graphene-ferroelectric hybrid memory devices.

The intrinsic, unusual electronic properties of graphene have made it a promising material for developing carbon-electronics,[1] such as RF-transistors,[2] and spin transport devices.[3,4] Additional functionalities may be introduced by integrating graphene with functional oxides such as ferroelectric Pb(Zr,Ti)O$_3$ (PZT), which can offer high-efficiency gating, local density modulation and memory functions through a spontaneous polarization that is large, electric field switchable and non-volatile. Integration of ferroelectrics with nanomaterials has previously been explored in nanowires[5] and carbon nanotubes[6] using crystalline films, and recently in graphene using a polymer layer[7] as the gate dielectric. In a previous study, we demonstrated high-efficiency carrier injection and superb electron mobility up to 140,000 cm$^2$/Vs at low temperature in graphene transistors fabricated on high-quality epitaxial PZT films.[8]

In this study, we report the observation of unusual resistance hysteresis in PZT-gated high mobility *n*-layer graphene (*n*-LG) field effect transistors (FETs). The metastable resistance states exhibit relaxation times of ~6 hours at 300 K and 80 days at 77 K, which is promising for constructing graphene-ferroelectric hybrid memories. The direction of the hysteresis and its thermally activated relaxation point to slow dynamics of graphene/PZT interfacial adsorbates as its origin.

We grow highly crystalline and smooth epitaxial PZT films on Nb-doped single-crystal SrTiO$_3$ substrates.[9] Characterization details are given in the supplementary material.[10] Piezoresponse force microscopy experiments show the as-grown polarization of the film, *P*, points uniformly towards the substrate. To reverse *P*, a negative voltage of 8V relative to the substrate is required.[10,11] Graphene sheets are mechanically exfoliated on PZT and identified optically (Fig. 1(a) inset). Thin sheets are examined with atomic force microscopy (AFM) and Raman spectroscopy. Figure 1(a) shows the AFM image of a multi-layer sheet, with the parts of 1, 2, 3 and 4 layers labeled. Raman spectra on the same sheet show clear G and 2D bands with a peak intensity ratio *I*(2D)/*I*(G) = 1.3 for single layers and 0.65 for bilayers. The 2D band becomes broader and progressively weaker in thicker flakes. Stand-alone thin flakes (*n*=2-15) are fabricated into FETs using standard lithography procedures.[8] Transport measurements are performed using lock-in techniques in a pumped $^4$He cryostat. The carrier density *n* is determined via Hall measurements, from which we deduce the gating efficiency of the device and the corresponding dielectric constant $\kappa$ of the PZT.[8] For devices reported here, $\kappa$ varies from 30-100.

In Fig. 2, we show the sheet resistance $\rho$ as a function of the backgate voltage $V_g$ of a 7-LG device at 300 K. The gating efficiency of this device is $1.35 \times 10^{12}$ cm$^{-2}$/$V_g$(V), corresponding to $\kappa \approx 100$. $\rho(V_g)$ exhibits distinct behaviors at low and high $V_g$. When the gate sweep is limited to $|V_g| < 2$ V (black curve), the carrier density and $\rho(V_g)$ follow conventional field effect modulation, and the forward and backward sweeps reproduce one another. $\rho(V_g)$ reaches a maximum at the charge neutrality point at $V_g = 0.17$ V. In this regime, we observe a Hall mobility $\mu_H$ up to 70,000 cm$^2$/Vs.[8] The small initial doping of $\sim 2 \times 10^{11}$/cm$^2$ indicates that the large remnant polarization of PZT has been mostly screened by charged adsorbates and surface reconstruction prior to graphene exfoliation.[12,13]

At $V_g > 2$ V, $\rho(V_g)$ becomes hysteretic, with the backward sweeping curve shifted to the right of the forward sweeping curve (Fig. 2). The onset of the hysteresis is accompanied by a saturation in $\rho(V_g)$, and a similar trend is observed in $n(V_g)$. The onset $V_g$ of the hysteresis upon forward sweep roughly coincides with the charge neutrality point in backward sweep, and vice versa. The shifts between the forward and backward $\rho(V_g)$ correspond to $\Delta n = 2.7 \times 10^{12}$/cm$^2$. Similar hysteresis is observed on four PZT-gated $n$-LG devices fabricated on three different PZT films,[10] regardless of the thickness of the graphene sheet (2-15 layers), its carrier mobility (16,000-140,000 cm$^2$/Vs) and the dielectric constant (30-100) of the PZT.

Several features of the hysteresis rule out ferroelectric polarization switching as its origin. First, the direction of the hysteresis is opposite to that expected from carrier density change induced by the polarization reversal of PZT. We thus refer to this behavior as "anti-hysteresis". Second, $\Delta n$ is only one percent of the nominal 2D charge density corresponding to the polarization of PZT ($\sim 3 \times 10^{14}$/cm$^2$). Third, the hysteresis occurs at $V_g$ smaller than the coercive voltage of PZT necessary to reverse the polarization.

The observed anti-hysteresis in $\rho(V_g)$ is reproducible and characterized by long relaxation times. In Fig. 2, the black curves are the stable state at lower $V_g$, while the red curves are more stable at higher $V_g$. If we sweep $V_g$ from 0 V following the black curve and pause at $V_g$ = 2 V, the system slowly relaxes to the red curve resistance $\rho_{red}$. The normalized resistance difference $\delta\rho(t)/\delta\rho(0)$ follows an exponential time dependence given by $\exp(-t/\tau)$ as shown in Fig. 3(a). Here $\delta\rho=\rho(t)-\rho_{red}$ is the time-dependent $\rho(t)$ relative to $\rho_{red}$ at $V_g$ = 2 V, and $\tau$ the relaxation time constant. $\tau$ increases from 6 hours at 300 K to 80 days at 77 K (Fig. 3(b)). The $T$-dependence of the relaxation rate, $1/\tau$, can be roughly described by:

$$\frac{1}{\tau} \sim \exp\left[-\frac{\Delta E_b}{k_B T}\right] \qquad (1),$$

suggesting a thermally activated relaxation process between two metastable states separated by an activation barrier. Figure 3(b) shows the Arrhenius plot of $1/\tau$ in two devices. The barrier height $\Delta E_b$ is estimated to be 50-110 meV by fitting to Eq. (1).

Such anti-hysteresis behavior was previously observed in carbon nanotube FETs gated by $SiO_2$ and epitaxial ferroelectric $BaTiO_3$ films.[6,14,15] Several explanations involving oxide charge trapping, surface charge trapping via silanol groups or the polarization of surface bound water molecules have been proposed.[16-18] Although the exact mechanism has yet to be clarified, studies in Ref. 15 point to a crucial role played by interfacial water molecules.

In our PZT-gated devices, charged adsorbates are present on the surface of PZT to screen its polarization prior to the exfoliation of graphene. As a likely candidate in the ambient, water is known to have two metastable forms when chemisorbed on the surface of transition metal oxides, such as $TiO_2$, $VO_2$ and $BaTiO_3$. It either maintains its molecular form or, in the presence of defects or uncompensated charges, dissociates into $H^+$ and $OH^-$.[19-21] The balance between the dissociation and recombination processes:

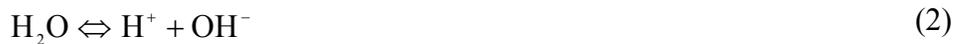

$$H_2O \Leftrightarrow H^+ + OH^- \qquad (2)$$

is influenced by the geometry of the lattice and external electric fields.[20,21] It has also been shown that $OH^-$ chemisorbs on the $Pb^{2+}$ sublattice of ultra thin $PbTiO_3$ to screen its polarization, with a binding energy of ~200 meV.[12]

Given the similarity between our observations and those reported for carbon nanotube devices, and the surface dynamics of water adsorbed on transition metal oxides, we propose to use the dissociation-recombination of water molecules adsorbed at the graphene-PZT interface to account for the anti-hysteresis seen in our devices. In this model, $OH^-$ and $H^+$ dissociated from chemisorbed water, as well as molecular $H_2O$, are likely to be present on the PZT surface to screen its polarization prior to the exfoliation of graphene. A positive $V_g$ applied to PZT reduces $P$ (originally uniformly polarized) and results in over-screening from the adsorbates. The balance of Eq. (2) is thus driven towards recombination. This process is thermally activated (Eq. (1)) due to the formation of a hydrogen bond between the $H^+$ ion and the surface anion $O^{2-}$, which acts as the activation barrier. The estimated barrier height $\Delta E_b \sim$ 50-110 meV is consistent with the binding energy of a weak hydrogen bond on a transition metal oxide surface.[12,20,21] Conversely, as $V_g$ decreases, $P$ increases and results in under-screening from the adsorbates. Dissociation of water is then favored to provide additional screening, which also needs to overcome $\Delta E_b$ on the order of a hydrogen bond. Both processes attempt to screen the polarization change of PZT, which in turn prevent charge injection into graphene and lead to the observed anti-hysteresis in resistance and carrier density. This model naturally explains the remarkable similarity among devices fabricated on films with a wide range of $\kappa$, since the relevant energy scales are determined by the surface chemistry of water on PZT, rather than PZT's bulk properties.

This robust hysteresis can potentially be used to implement a memory device, where the "0" and "1" states are represented by the low (red) and high (black) resistance states, as shown in Fig. 2. The high/low resistance ratio is 2 for the 7-LG and reaches 3.5 in a 2-LG device.[10] The high-$\kappa$ nature of PZT (up to 100) and the superb room temperature carrier mobility in these devices (up to 70,000 cm$^2$/Vs) should enable reading and writing operations at small gate voltages and exceedingly high frequencies. To achieve the retention time necessary for a practical non-volatile memory device, controlled adsorbates with higher binding energies may be required.

In conclusion, we observe an unusual resistance hysteresis in high-mobility $n$-layer graphene FETs fabricated on epitaxial PZT thin films. We attribute its origin to the dynamic dissociation/recombination of water molecules chemisorbed at the PZT surface. The reproducibility of the hysteresis, combined with excellent carrier mobility, provide a viable pathway of implementing memory functions in graphene-oxide hybrid devices.

We are grateful to V. Crespi, P. Eklund, V. Henrich, P. Paruch, and X. Pan for helpful discussions and S.-H. Cheng and B. Wang for technical assistance. Work at Penn State is supported by NSF NIRT No. ECS-0609243 and NSF CAREER No. DMR-0748604. Fabrication of samples at Yale is supported by NSF MRSEC No. DMR-0520495, NSF No. DMR-0705799, ONR, and NRI. The authors also acknowledge use of NSF NNIN facilities at PSU.

**Figure caption:**

Fig. 1 (a) AFM image of a multi-layer graphene sheet on a 300 nm PZT film. The layer numbers are determined by height measurements and marked in the figure. Inset: optical image of the same sheet. Scale bar: 10 μm. (b) Raman spectra on different parts of the sheet shown in (a) normalized to the G peak intensity. The increasing background at low wave numbers and the small broad peak centered at 1615 cm$^{-1}$ are from the PZT substrate.

Fig. 2 $\rho(V_g)$ on a 7-LG FET at 300 K. $\rho(V_g)$ reaches a maximum at the charge neutrality point. Arrows indicate the sweeping direction of $V_g$. The top axis labels the corresponding carrier density scale. ($V_g$ = 2 V marked by the blue triangle.) The "0" and "1" labels illustrate the "off" and "on" states for a memory operation.

Fig. 3 (a) Time-dependent relaxation from the metastable resistance state (the black curve in Fig. 2) towards the stable state (the red curve) at $V_g$ = 2 V. The blue dashed line is an exponential fit to exp($-t/\tau$) with $\tau$ = 6 hour. (b) Arrhenius plot of $1/\tau$ vs. temperature from measurements in (a) (solid squares) and similar measurements on another 15-LG device (open triangles). Blue dashed lines are fittings to Eq. (1) with $E_b$ = 50 (upper line) and 110 meV (lower line).

Figure 1

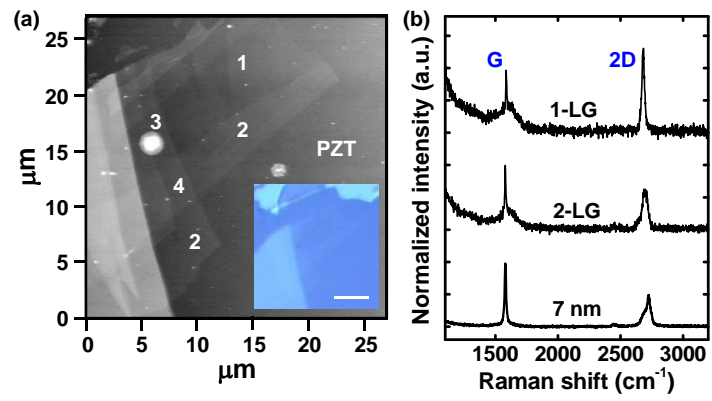

# Figure 2

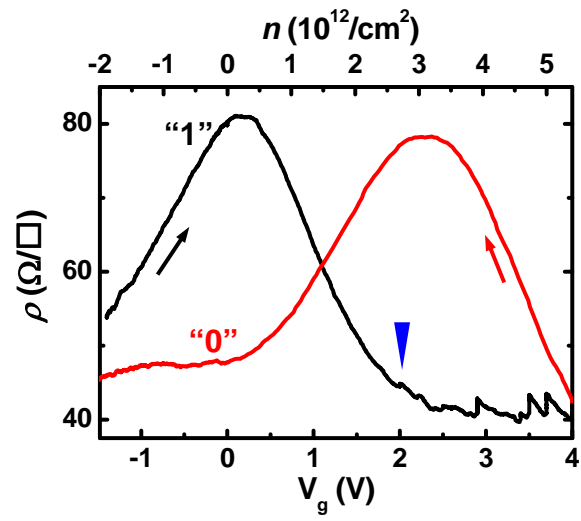

Figure 3

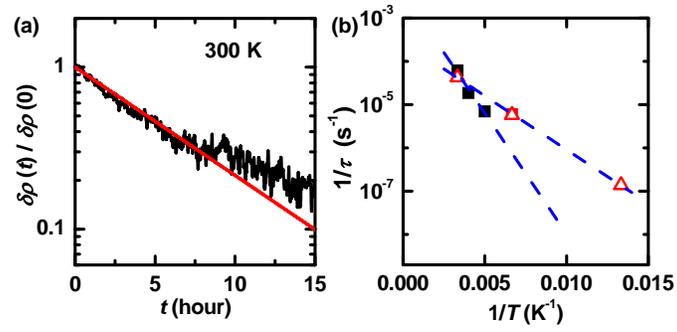

**Unusual Resistance Hysteresis in *n*-Layer Graphene Field Effect Transistors Fabricated on Ferroelectric Pb(Zr$_{0.2}$Ti$_{0.8}$)O$_3$**

X. Hong,[1] J. Hoffman,[2] A. Posadas,[2] K. Zou,[1] C. H. Ahn,[2] J. Zhu[1]

[1] Department of Physics, The Pennsylvania State University, University Park, PA 16802

[2] Department of Applied Physics, Yale University, New Haven, CT 06520

**Supplementary Information Content:**

1. Characterization of Pb(Zr$_{0.2}$Ti$_{0.8}$)O$_3$ films.

2. Anti-hysteresis observed on a 2-layer graphene field effect transistor.

1. **Characterization of Pb(Zr$_{0.2}$Ti$_{0.8}$)O$_3$ films**

Pb(Zr$_{0.2}$Ti$_{0.8}$)O$_3$ (PZT) films used in this study are epitaxially grown on Nb-doped single-crystal SrTiO$_3$ (STO) substrates via radio-frequency magnetron sputtering. Atomic force microscopy (AFM) measurements show a typical surface roughness of 3-4 Å. X-ray diffraction (XRD) measurements show a predominantly $c$-axis oriented film growth. The $c$-axis lattice constant is ~4.15 Å, consistent with the value of a partially relaxed film. The typical rocking curve of the 001 peak has a full-width-half-maximum of 0.1-0.4°, reflecting a high degree of crystallinity.[1]

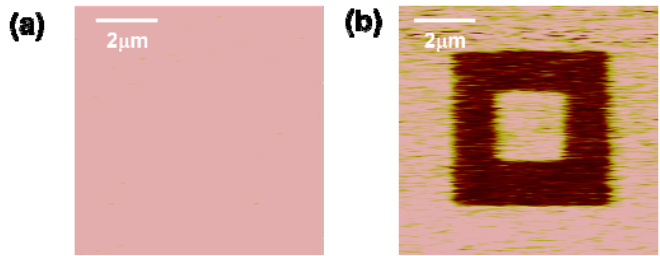

Fig. S1 (a) PFM measurement taken on an as-grown 400 nm PZT film. (b) PFM image of the same area in (a) after two AFM writings. An area of (5 μm)$^2$ in the center is written with $V_{tip}$ = -12 V first. The (2 μm)$^2$ inner square is subsequently re-written with $V_{tip}$ = +12 V. The signal contrast indicates the polarization of the dark area is opposite to that of the as-grown film. The image is taken 17 hours after writing.

We examine the polarization $P$ of the PZT film with AFM writing and piezoresponse force microscopy (PFM). A biased AFM tip scans in contact with the PZT film to polarize the ferroelectric domain underneath.[1] The tip bias $V_{tip}$ varies in polarity and magnitude. The resulting domain structure is then imaged with PFM, where the phase of the piezoelectric signal sensitively depends on the direction of $P$.[1] Figures S1(a) and S1(b) show the PFM image of a 400 nm PZT film before and after an AFM writing procedure. Before the writing, the PFM image of the as-grown film displays a spatially uniform signal, indicating a uniform polarization across the whole film (Fig. S1(a)). Figure S1(b) shows the PFM image of the same area after a (5 μm)$^2$ square area is written with $V_{tip}$ = -12 V and subsequently re-written with $V_{tip}$ = +12 V in just the (2 μm)$^2$ center area (bright). We observe a strong contrast between the background (outer area) and the (5 μm)$^2$ area written with $V_{tip}$= -12 V (dark), and essentially no contrast between the background and the $(2~\mu m)^2$ inner square re-written with $V_{tip}= +12$ V (bright). The image shown in Fig. S1(b), taken 17 hours after writing, shows no appreciable change compared with another image taken immediately after writing. This observation demonstrates that the PFM signal is indeed due to ferroelectric switching instead of spurious charging effect. We thus conclude that the polarization of the as-grown film $P$ points uniformly into the surface, as expected from the growth procedure.[1] To reverse the polarization of a freshly polarized area, a negative voltage pulse (100 ms) with a minimum magnitude of 8 V is necessary.

## 2. Anti-hysteresis observed on a 2-layer graphene field effect transistor

Figure S2 shows $R(V_g)$ of a 2-layer graphene (2-LG) device fabricated on a 300 nm PZT film. The gating efficiency of the film is $0.49 \times 10^{12}$ cm$^{-2}$/ $V_g$(V), corresponding to a PZT dielectric constant of $\kappa \approx 30$.

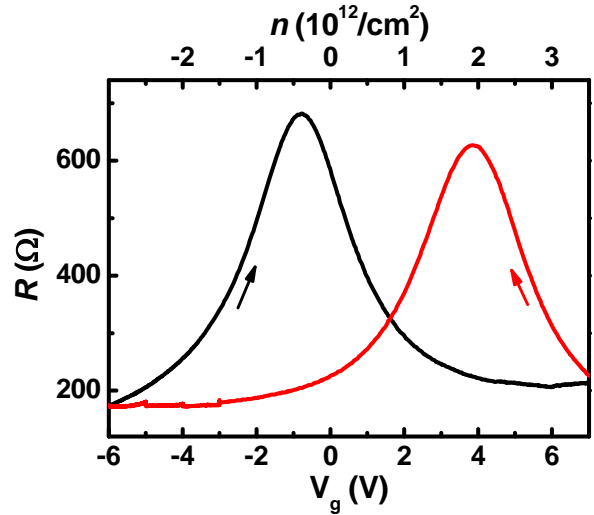

Fig. S2 $R(V_g)$ taken on a 2-LG FET at 100 K. Arrows indicate the sweeping direction of $V_g$. The top axis labels the corresponding scale of carrier density modulation. The maximum in $R(V_g)$ corresponds to the charge neutrality point.

The $R(V_g)$ of the 2-LG exhibits similar anti-hysteresis as the 7-LG device reported in the main text. The shift between the forward and backward gate sweeps corresponds to $\Delta n = 2.4 \times 10^{12}$/cm$^2$, comparable to that of the 7-LG device ($\Delta n = 2.7 \times 10^{12}$/cm$^2$). The maximum high/low resistance ratio is 3.5 in this device.

The mobility of the 2-LG device in the low $V_g$ field effect modulation regime exhibits $\mu$ = 16,000 cm$^2$/Vs at 100 K. The mobilities of PZT-gated $n$-LG ($n$ = 2-15) devices vary considerably from 16,000 cm$^2$/Vs to 140,000 cm$^2$/Vs.[2] We surmise that the density of screening adsorbates and their spatial distribution are important in determining $\mu$, since lower mobility devices tend to show higher initial doping, suggesting incomplete screening of the PZT film. It is also apparent that given the amount of scattering caused by charged adsorbates at the PZT surface is large[3,4], the dielectric screening of the PZT film plays a key role in achieving the mobility values observed here.